# A Survey on Security Issues in Modern Implantable Devices: Solutions and Future Issues

Emmanuel Kwarteng, Mumin Cebe
Computer Science Department, Marquette University
emmanuel.kwarteng@marquette.edu, mumin.cebe@marquette.edu

*Abstract*—**Implantable Medical Devices (IMD) is a fast pace growing medical field and continues to grow in the foreseeable future. Advancement in science and technology has led to the IMD devices offering advanced medical treatments. Modern IMDs can automatically monitor and manage different patients' health conditions without any manual intervention from medical professionals. While IMDs are also becoming more connected to enhance the delivery of care remotely and provide the means for both patients and physicians to adjust therapy at the comfort of their homes, it also increases security-related concerns. Adversaries could take advantage and exploit device vulnerabilities to manipulate device settings remotely from anywhere around the world. This manuscript reviews the current threats, security goals, and proposed solutions by comparing them with their strengths and limitations. We also highlight the emerging IMD technologies and innovative ideas for new designs and implementations to improve the security of IMDs. Finally, we conclude the article with future research directions toward securing IMD systems to light the way for researchers.**

## I. INTRODUCTION

Healthcare is a massive industry in the US that contains over 750,000 companies and consumes over 15% of gross domestic product (GDP) [1]. Nowadays, this industry is on the cusp of a technological revolution by the rapid integration of smart devices to create a digital transformation in healthcare delivery [2]–[4]. The healthcare industry has started to offer new services from digital health records to remote e-health services and is expected to increase these services across multiple channels via remote access smart medical devices and IoT-based health devices.

In this context, the healthcare domain is also experiencing numerous advancements in Implantable Medical Devices (IMDs) domain [5]. The integration of computing with IMDs has changed the landscape of modern medicine. We have been witnessing the explosion of implantable device applications with the improved techniques to manufacture low cost integrated circuits that provide real-time monitoring and treatment to check a patient's health status. Low-power wireless connectivity [6] and development of numerous lightweight communication protocols [7], [8] have helped to make small-scale systems that does not require big batteries. These systems can collect a range of physiological values like blood pressure, heart rate, temperature, oxygen saturation, and neural activity, and can help medical field personnel to identify appropriate treatments. As a result, nowadays, there are various IMDs that remotely collect a patient's physiological data to provide automated treatment. For instance, an implanted glucose monitor can automatically regulate sugar levels by injecting insulin into a patient and its wireless capability allows to transfer data for monitoring and getting adjustments related to treatment configuration without sacrificing patient mobility or physically accessing the device [9].

The global medical device market is expected to reach an estimated $409.5 billion by 2025, and it is forecast to grow at a Compound Annual Growth Rate (CAGR) of 4.5% from 2018 to 2023 [10]. At the same time, the functional complexity of the medical devices is increasing day by day, in parallel with the development of various energy-efficient communication protocols [7] that together boosts the application of IMDs since collecting physiological values of a patient are becoming more and more efficient. These IMDs are gaining popularity in United States and non United States market.

Nonetheless, the expected digital revolution in healthcare introduces worrisome cybersecurity risks, considering the expanded attack surface with the diversity devices in healthcare systems. Any cyberattack to the evolving digital healthcare systems can easily cause massive disruptions in healthcare delivery. The threat is authentic and getting worse in tandem with the pace of digitization of healthcare. Even though integrating cybersecurity-related countermeasures is increasingly becoming a critical part of any healthcare organization, the current situation is far behind the urgent needs. This explains the jaw-dropping 90% ratio among healthcare companies who have experienced at least one security breach incident in the last two years [11]–[13].

Overall, IMDs are becoming an integrated part of the healthcare ecosystem and their increased access capability paves the way for the attackers to exploit the security and privacy. The concerns even caused backstepping where doctors disables the wireless connectivity of IMDs that is implanted to some critical politicians to protect it from being hacked [14]. Besides, researchers found several vulnerabilities on commercial IMDs, as a result, they are also trying to identify the attack vector of IMD enabled healthcare systems [5], [15]–[18]. There are also solutions to address some of these attacks [15], [19]–[21]. However, these studies are still in its infancy phase and security of IMDs needs more attention to build a secure ecosystem, before IMDs become more and more ubiquitous in modern healthcare systems.

**Contributions:** The following are the contributions of this manuscript to the IMD security highlighting its difference from related works in this field:

- We provide a general background and taxonomy of current IMD solutions in the market.
- We researched and analyzed IMD solutions to identify their implementation details to guide researchers while tackling security issues.
- We presented emerging IMD solutions, technologies, future trends of IMD, and advancement of the market in IMD space.
- We delivered a security analysis of domain by categorizing studies under cryptographic keys security, firmware update security, sensors and actuators security to serve as guidelines to researchers to identify related open problems.
- The firmware update section is particularly important and sets light to researcher while tackling with this recent aspect of IMD security.
- We evaluated current and most common sensor-based attacks and threats.
- We identified future research directions that needs immediate attention to improve the security and privacy in IMD space.

This manuscript is structured as follows: Section II opens up with a background information about IMD applications, communication protocol, data processing and emerging IMD technologies. Section III presents current security threats and attacks on IMD as well as the security solutions and goals. Section IV will cover Cryptographic key security which can be deployed to meet a security goal. Section V will also focus on another security goal by reviewing software and firmware update solutions. Section VI presents sensors and actuators security solutions. Section VII will focus of future research and development areas to improve security in IMD field. Section VIII presents the conclusions of this work.

## II. BACKGROUND

In this section we describe the general overview of Implantable Medical Device Solution as depicted in Fig. 1 and further define the key features used throughout the manuscript.

*1) Implantable Medical Devices:* A medical device is defined as implantable if it is either partly or totally introduced, surgically or medically into the human body and is intended to remain there after the procedure [22]. [23] also defines Implantable Medical Devices as electronic systems embedded in the human body to continuously monitor its health, detect and predict certain conditions and deliver therapies. As defined by [23], Implantable medical devices are purposely introduced surgically in the body of a patient to help solve underlying health problems of a patient or help to improve the functionality of a patient's body parts. The location where the IMD is implanted surgically depends on which health problem the physician is trying to solve. For example, a patient with unbearable back pain may consent to a spinal cord stimulation which is implanted into the epidural space while a patient with heart problems may consent to pacemaker implanted surgically near the heart. Another example is when a patient with problems such as Parkinson's or epilepsy will consent to a deep brain stimulation IMD implanted closer to the brain. These devices could operate as standalone medical devices or connect to a network of devices. Modern IMDs communicate with an external device called the programmer using Bluetooth Low Energy protocol as an example to send monitoring data or receive updated therapeutic regimens. In contrast, the previous generation of IMDs communicates with external programmers using a wireless communication protocol, which is restricted by distance, normally within 2m range of proximity [23]. IMDs can help manage a broad range of ailments including cardiac arrhythmia, diabetes, Parkinson's disease, epilepsy disease, back pain, and spinal cord diseases [24] to name a few. Researchers and implantable medical device manufacturers are advancing science and exploring IMDs to manage aches and pains in the human body. As of this writing, millions of Americans have undergone surgery to implant some type of medical device [25].

### A. Implantable Medical Applications

Fig. 2 shows the general overview of IMD solutions currently in the market. Implantable medical devices are unique in their applications, targeted therapy, and functionality. Therefore, no two IMDs can be compared easily with each other. For instance, a Cardiac implant delivering therapy to a patient with heart problems may focus on the cardiac system, while a Deep Brain Stimulation implant delivering therapy to a patient with Parkinson's disease may focus on the central nervous system. However, these two systems would have different impacts on what resources (e.g. communication protocols, authentication mechanisms) while implementing them. Details about some IMD solutions are described below:

- **Cardiac implant**: Cardiac implantable medical device is an IMD implanted under the skin of the patient near the upper chest, which monitors the electrical activity and applies electrical impulses of suitable intensity to make the heart pump regularly [26]. Cardioverter Defibrillators (ICD) and Pacemakers are the most common cardiac IMDs. An Implantable ICD system is comprised of a pulse generator, and one or more leads for pacing and defibrillation electrodes [27]. The leads also contain bipolar electrodes which are used for ventricular pacing and sensing [27]. These devices could automatically monitor and analyze cardiac rhythms and deliver defibrillation shocks when ventricular fibrillation is detected.
- **Deep Brian stimulation**: Deep Brain Stimulation (DBS) is an IMD used in functional neurosurgery to deliver continuous electrical stimulation through implanted electrodes [28]. The electrodes are implanted at a specific brain area and generate electrical impulses to reduce tremors, muscle stiffness, and other brain disorders. DBS is mainly used to offer therapy to patients with a movement disorder, Parkinson's disease, Essential Tremor, Dystonia, Tourette Syndrome, Depression, and Obsessive-Compulsive Disorder [29].
- **Spinal cord stimulation (SCS)**: Spinal cord stimulation [30] is implanted a few centimeters under a patient's

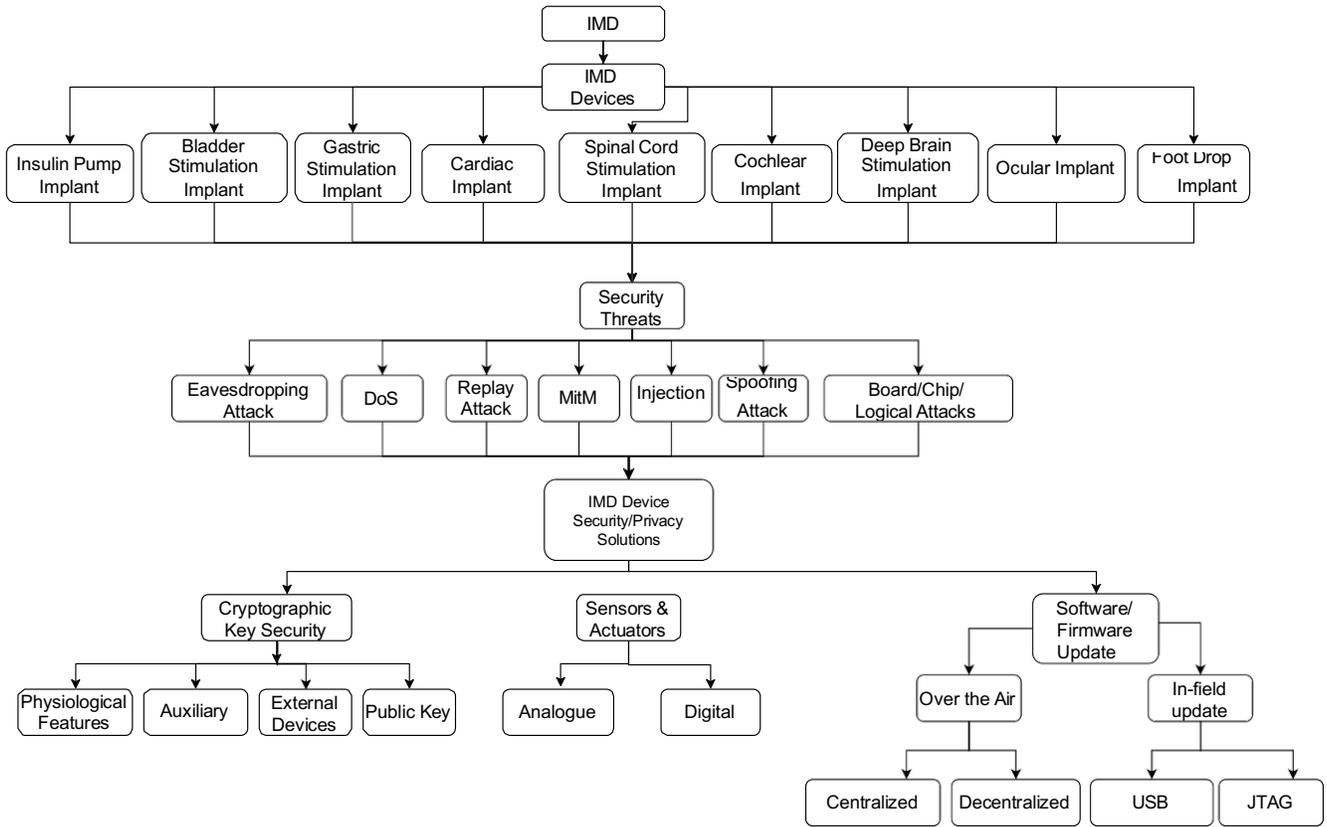

Fig. 1.  Implantable Medical Devices Taxonomy

skin, which sends manageable electrical pulses to the sensing pathway of pain in the spinal cord and interferes with the nerve impulses sent to the brain and replaces it with a tingling sensation. A remote controller can remotely control the stimulator to adjust stimulation based on the level of pain and daily activities.

- **Cochlear implant**: A cochlear implant [31] is a small electronic device that electrically stimulates the cochlear nerve. The cochlear implant has two parts (the outer part sits outside the ear, and the internal part is implanted under the skin behind the ear). The outer part processes the sound and transmits it to the internal part of the implant. The thin wire and electrode send signals to the cochlear nerve, which sends sound signals to the brain to produce a hearing sensation. This IMD is used to restore the hearing of patients who are suffering from hearing loss or deafness.

- **Bladder stimulation implant**: Bladder stimulation [32] is an IMD that focuses on generating sacral electrical neuromodulation for stimulating the bladder. This IMD is often used to treat overactive bladder, stress urinary incontinence, urinary retention, and urinary tract infection, common to a patient with spinal cord injury. This IMD comprises an outer part (controller, transmitter) and an implantable microstimulator (implant and electrode).

- **Foot-Drop Implant**: Foot-drop implant [33] is a two-channel implantable stimulator connected through a cable with a multipolar nerve cuff electrode which is implanted on the peroneal nerve proximal to the knee. This implant is powered and controlled by an external control unit. The device makes it possible to activate the dorsiflex muscles selectively to restore lost ankle function.

- **Gastric Stimulation**: Implantable Gastric stimulator (IGS) [34] is an IMD that is placed through the abdominal wall (laparoscopy) and has two implantable components. Like the pacemakers, the IGS system comprises the implantable Gastric Stimulator, the lead, and the programmer. The lead with two electrodes is implanted into the gastric lesser curvature muscle tunnel. The device is generally used to control gastric motor dysfunction to treat morbid obesity [35], severe gastroparesis [36], and severe obesity [34].

- **Implantable Wireless Pressure Sensor**: Implantable Wireless Pressure Sensor(IWPS) [37] is implanted closer to the organ, nerve or tissue that requires stimulation. This IMD has been deployed historically to treat multiple diseases of disorder. Pressure is an important physiological parameter in various organs and is considered as one

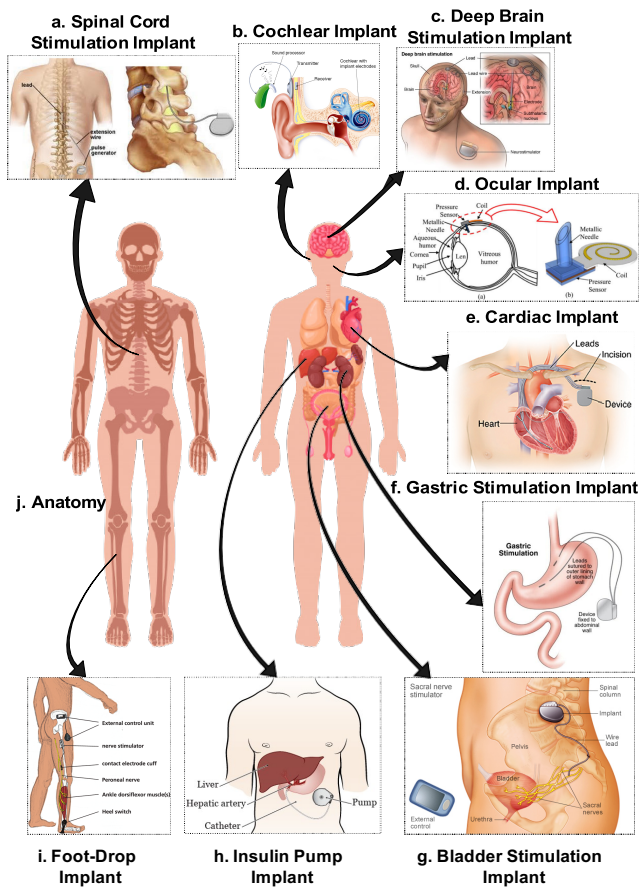

**a. Spinal Cord Stimulation Implant**

**b. Cochlear Implant**

**c. Deep Brain Stimulation Implant**

**d. Ocular Implant**

**e. Cardiac Implant**

**f. Gastric Stimulation Implant**

**j. Anatomy**

**i. Foot-Drop Implant**

**h. Insulin Pump Implant**

**g. Bladder Stimulation Implant**

Fig. 2. Implantable Medical Devices

of the key parameters for health and disease diagnosis. Wireless communication technology deployed in pressure monitoring has increased patient comfort and prevented infections from the catheter or wires used in older versions of pressure monitoring systems. As a result, the Federal Communication Commission (FCC) established Medical Implant Communication Service (MICS) [38] band such as 402-405 MHz to serve as a dedicated frequency for Implantable Devices. The following are sample areas where IWPS has been used to treat patients.

- **Intraocular Pressure Monitoring**: Intraocular Pressure Monitoring (IPM) [39] is implanted in the supertemporal region of the human eye between the superior and lateral rectus muscles and continuously monitors Intraocular Pressure(IOP). The IPM is used to treat Glaucoma disease which can lead to blindness [39].

- **Bladder Pressure Monitoring**: Bladder pressure monitoring (BPM) [37] is considered to be one of the key elements in diagnosing bladder dysfunction and consist of the base station, pressure sensor probes, and the antenna. The antenna is used to transfer pressure measurements of the bladder to the base station.

- **Cardiac Pressure Monitoring**: Cardiac Pressure Monitoring (CPM) [40] is used to remotely identify poten-

tial heart failure by monitoring the left ventricular filling pressures which is found to proceed hospitalization for heart failure.

- **Emerging IMD Applications**:

  – **Integrated Intelligence for Vascular Access**: Vascular access is a commonly used medical technique to make it easy to give injections or draw blood for labs. It is basically a long-term port placed into a vein for patients having chemotherapy over a long period of time [41]. There are emerging applications that aim to re-purpose this technique in which an implantable device that leverages advanced sensors and communication is integrated into the vascular access port [42]. With the help of this integrated sensing and communication capability, patients can continuously and remotely be monitored that helps to identify the related treatment-related complications remotely.

  – **Gum Health Monitoring**: It is a new technology that leverages special sensors and wireless communication to obtain the health of gum and teeth to monitor patients' entire oral health [43]. Even though the current version is not exactly implantable, it is not hard to imagine implantable versions will emerge soon to diagnose gum disease, gum recession, plaque level, etc,. It may even also follow saliva to identify harmful pathogens, such as bacteria or viruses, that gives clues about different diseases [44].

  – **Energy Generation**: All implantable medical devices require some amount of energy from a power source. Most of the IMDs derive its source of energy from batteries which deplete overtime. Most non rechargeable battery has between 10 and 15 years of life. At the end of life of the batteries, patients go through a surgical procedure to replace the battery. To advance this generation of energy for IMDs, researchers are investigating the possibilities of harvesting power from organisms (e.g heartbeat and respiration) or surrounding environment [45]. Another area of emerging energy generation IMD technology is self-powered IMD devices.

### B. Communication Protocols

Connectivity is becoming a crucial part of IMD solutions. Patient with implantable medical devices needs to be monitored and therapy adjusted in and out of clinicians office. Therapy delivery and monitoring of patients with implantable medical devices both remotely and in the healthcare facility is made possible by the communication protocols such as BLE, WIFI, Zigbee, and Z-Wave as shown in Fig. 3. Besides, there are IMD devices that use a regulated radio band named Wireless Medical Telemetry Service (WMTS). The Federal Communications Commission (FCC) established it by allocating specific frequency bands exclusively for wireless medical devices. The WMTS has 14 MHz of spectrum in three defined frequency bands of 608-614 MHz, 1395-1400 MHz, and 1427-1432 MHz for primary or co-primary use by eligible wireless medical telemetry users [46]. In addition to the listed radio frequency-based technologies, there are IMDs that uses

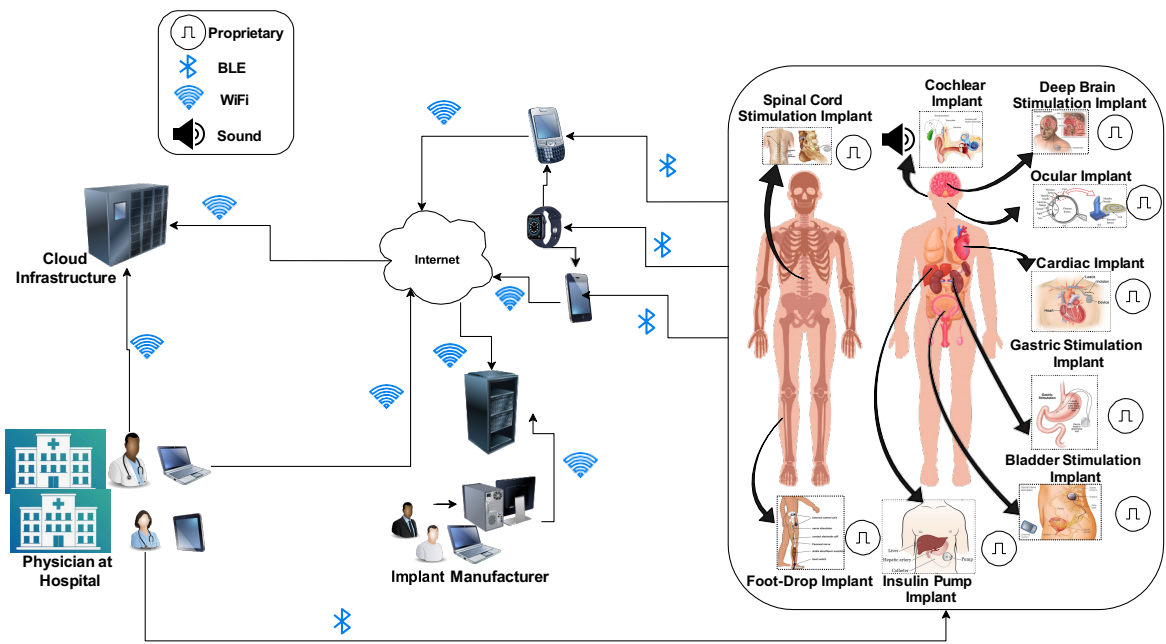

Fig. 3. IMDs communication

"sound" as a communication channel and data is transferred using sound waves with special modulators and demodulators [47].

### C. Data and Data Processing

Data processing is the process that information collected from patients, devices and interested parties goes through before they can be used for the intended purpose. Data processing provides the opportunity for data analysis, use, analytic and research with patient consent. It is used to enhance therapy, diagnose illnesses, and facilitate treatment and pain management which are crucial function of IMDs. For instance, an IMD can be programmed to deliver the right amount of signal to the brain to stimulate the brain cells to a patient with movement disorder. Depending on the intended use of the data, these data might go through different stages from its collection, use, transfer, and storage.

- **Data Collection**: The process of data collection in IMDs is totally dependent on the type of IMD. For instance, neural signals need pre-processing for noise removal before even start transferring. In addition, data that is collected with the organization's trust zone (behind the firewall) might not need a rigorous data validation as compared to data collected from a non-trust zone (over the internet).
- **Data Transfer**: The collected data from the IMDs are transmitted to the destination using different communi-

cation protocols. During transfer, the integrity, confidentiality, and authentication of it is ensured using various cryptographic methods.
- **Data at Rest**: After data is collected, and transferred, it is stored in a location such as databases, clouds or the organization's back-end infrastructure.

## III. RELATED WORK

In recent years, a few surveys have been carried out reviewing and reporting on security and privacy problems and solutions in healthcare domain that also covers some IMD related concerns. This section is used to highlight recent surveys and discuss how the differ from this survey.

Newaz et al. extensively reviewed security and privacy of overall healthcare systems including devices, sensors, networks, communication, healthcare providers and also compares attacks and solutions [48]. Hathaliya et al. provided an overview of different application areas of healthcare 4.0 and discussed the integration of different technologies by comparing architecture, research questions, open issues, research challenges, security algorithms and taxonomy of recent surveys in healthcare industry [49]. Whipple et al. researched the aspects of firmware on embedded systems including reverse engineering, and different ways to retrieve firmware binaries from an embedded device [50]. Wu et al. presented a comprehensive survey on classification of access control schemes and focusing specifically on security incidents, threat model, and regulations for IMD security [51]. Aram et al. studied IMD architecture, network interface and communication protocol vulnerabilities while also comparing security and networking of medical body area network to wireless sensor network. Their work also reviewed conflicting requirement of IMD design and security needs. McGowan et al. summarized known vulnerabilities landscape (security issues, unpatched devices, authentication), cyberattack landscape (cardiac, neuromodulation, implantable mobile devices) of medical Internet of Things (mIoT) devices and the resulted patient safety [52]. Ameer et al. reported the main security goals of next generation IMD and relevant protection by reviewing advent events and different monitoring systems. Their survey also compared recent technologies used, their algorithm if any, the databases used and the limitations of these technologies [53]. Sikder et al. explored current threats against sensors in IoT devices, existing countermeasures, and research approach in addressing sensor security. They also presented flaws and mitigation of sensor management systems, by presenting a threat to sensors (information leakage, transferring malware, false data injection and denial of service) and the corresponding existing sensors research work [54]. Sun et al. studied security and privacy challenges, requirements, threats, and future research directions in Internet of Medical Things (IoMT). Their study specifically looked at systems and network design challenges, security and privacy requirement at the data, sensor, personal and medical server levels as well as research in these areas [55]. Giraldo et al. provided an overview of security and privacy in cyber physical system (CPS) which includes medical devices, manufacturing, industrial control systems, intelligent transportation systems, and grid systems. Their study compared surveys with concentration on security, privacy and defense of CPS [56]. Razaque et al. presented a survey on recent cyber-attacks in medical field, classified these attacks based on the medical systems weaknesses and reviewed strength and limitations of previously recommended solutions. They also reported on countermeasures for weak cyber security architecture and recommendation for future cyber security research [57]. Malamas et al. analyzed and compared current IoMT risk assessment methodologies to identify common theme and implementation gaps. Based on their analysis, they provided security controls that can be used to mitigate those risks [58]. Koutras et al. surveyed and classified IoT communication protocols based on their applicability to IoMT domain while examining security characteristics and limitation of these protocols. Their survey also identified mitigating controls to current attacks on IoMT communication protocols [59]. Oh et al. conducted a comprehensive survey on security and privacy issues, security concerns, requirement and solutions with e-health data, medical devices, medical networks, edge, fog and cloud computing. The survey also discovered research trends and challenges for e-health systems [60].

Although these surveys cover various cyber-attacks, vulnerabilities and some countermeasures in IMD domain, they are clearly lack of particular focus which is consolidated security solutions in IMD domain, and their focus is on rather IoT, IoMT, CPS, mIoT and healthcare 4.0. This survey is dedicated to IMD and IMD related studies. Our main goal is categorizing the studies while conveying some insights from other domains to inspire researchers to address yet not-solved security problems of IMD.

## IV. Security in IMD Domain

Cybersecurity is one of the pillars of the current digital transformation and is defined as the state where information and systems are protected from unauthorized activities, such as access, use, disclosure, disruption, modification, or destruction [61]. Its basics are modelled as confidentiality, integrity, and availability, known as "CIA triad".

As threats are increasing and evolving, cybersecurity becomes one of the most important public safety concerns that even led to an executive order about it [62]. Considering the crucial nature of IMDs and their services to patients, ensuring security is becoming the highest priority of IMD manufacturers and healthcare providers. In addition to that, IMDs have an increased attack surface due to its requirement scheme that accepts many parties (i.e., programmers, administrators, patients, and physicians) as authoritative actors. This wide attack surface already carries many unique challenges to address; besides, recent ransomware attacks in the healthcare industry have created additional potential risks for IMDs [63]. Some common threats that pose a challenge to IMD security as shown in Fig. 1, includes but not limited to the following:

### A. Security Threats

- **Eavesdropping Attack**: IMDs which communicate over insecure channels are vulnerable to eavesdropping attack. For instance, an attacker close to the patient could read/modify messages exchanged between IMD and controller to determine the commands, the type of IMD, serial number, and manufacturer. Similarly, a remote attacker who compromised the healthcare facility network can capture and modify readings of all IMDs within the facility or even the readings from the ones that are outside of the facility, which allows remote data collection..
- **Denial of Service (DoS) Attack**: An attacker could launch an attack on IMD to drain the battery life of the IMD which will require a surgical procedure to replace the batteries. An attacker could submit bogus requests to the IMD, which will cause the IMD to process these requests, therefore creating a computational overhead which eventually drains the battery. An attacker could also block or jam the IMD frequency band to launch a denial-of-service attack. A malicious attacker could also deny service when a successful malware executed on the IMD causes components on the IMD to overheat and shutdown.
- **Replay Attack**: An attacker can intercept and resubmit the same request from a legitimate controller/programmer that has already established trust with the IMD to alter its state.
- **Man in the Middle Attack**: A MitM attacker who successfully place himself/herself between the IMD and the controller or between the controller and the remote server can play an active role to intercept patient's private data, recover authentication tokens, store previously executed commands, and replay or block commands.

- **Software Injection Attack**: A successful injection attacker on the IMD could lead to an attacker replacing the firmware on the IMD to gain full access or injecting their own code into the flash or memory to force the IMD to execute unauthorized commands to adjust patient's therapy.

- **Side Channel Attack**: Side-channel attacks differ significantly from the previously discussed attack types. They leverage information such as timing, power consumption, electromagnetic and acoustic emissions to steal data. Considering a typical IMD comes with various sensors, side-channel attacks pose a vital threat against IMD security by enabling unconventional data leakage scenarios.

### B. Security Goals

Due to security challenges discussed above, security goals must be established at the onset of a new feature development, improvement to existing IMDs, new IMD development and considerable evaluation of legacy IMDs in the field to ensure the necessary counter measured are put in place to limit the success of the attacker. CIA triad is the fundamental of information security which becomes the goal of all IMDs.

- **Confidentiality in IMDs**: Communication between IMDs, programmers, remote controls, hospital infrastructure and/or IMD manufacturer often requires exchange of sensitive information that needs to be protected from malicious actors. Sensitive data such as patient identifiable information, health records, IMD state and usages, IMD monitoring information, IMD audit information and user behavior related data need to be protected from unauthorized users including hackers, internal and external malicious actors. Data should be kept confidential during transfer and storage. Due to the nature of IMDs and functionalities, this can lead to vulnerabilities which can compromise the confidentiality of sensitive data.

- **Integrity in IMDs**: The integrity of information between IMDs and programmer or remote servers should be ensured, and the source of the data should be verified. The correctness of data is key to offering the right therapy patients require for their treatment. Thus, the integrity of the IMD's data defines the quality and effectiveness of healthcare patients receive for their treatment.

- **Availability in IMDs**: The IMD solution is expected to offer continuous stimulation or monitoring of the target treatment. IMD should be made available to provide therapy to patients and available when physicians need access to adjust stimulation. IMDs should be resistant enough against denial-of-service attacks to prevent depletion of their resources.

## V. Cryptographic Key Security

As in the case of conventional systems, the listed security goals in the previous section can be met by using cryptographic keys as depicted in Fig. 1. However, the security of the keys is a major issue and involves ensuring the security of key management operations such as generation, storage, distribution, use, maintenance, and revocation of cryptographic keys. These operations are applied under two different setups: symmetric key and asymmetric key.

Symmetric key setup is mostly utilized to encrypt data and undoing this is fundamentally difficult without knowledge of the key itself. This setup is called "symmetric" since the parties utilize the same secret key for a cryptographic operation such as encrypting and decrypting the data. Symmetric keys are commonly used to provide data confidentiality by encryption& decryption services and like digital signatures, to ensure the origin and integrity of data through message authentication codes (MACs).

Asymmetric key setup, generally known as public-key Cryptography (PKC), utilizes a key pair (e.g., a public key and private key) to perform cryptographic operations. Anyone can know the public key, but its pair, the private key, is just known by the party who creates this key pair and should be secret. In PKC, the particular key of the key pair is used for different purposes to provide security, and their usage depends on the cryptographic service to be provided. However, they are mostly used to ensure the origin, identity, and integrity of the data through digital signatures and distribution of symmetric keys between parties.

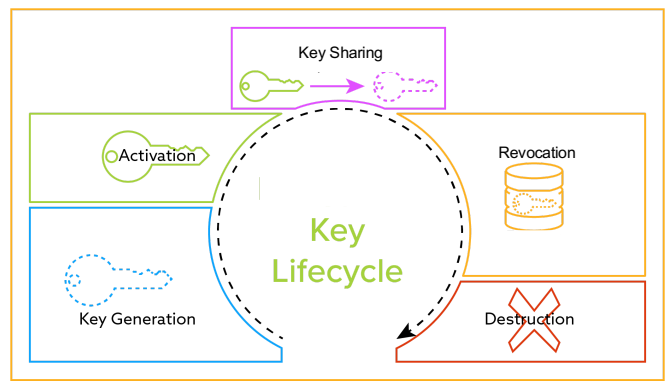

Fig. 4. The life-cycle of cryptographic keys

Considering the security of cryptographic operations highly depends on the secrecy of the used key material, assuring the safety of the keys by a proper management of key-life cycle is essential for security. The following subsections introduces the studies in IMD domain to govern the key life cycle as shown in Figure 4.

### A. Employing Physiological Features (PF)

Random numbers are essential when generating cryptographic keys and nonces to establish secure cryptographic systems. However, IMDs' microcontroller-based architecture does not have a traditional true randomness source to generate random numbers. Thus, there are studies that offer the use of physiological signals such as ECG, photoplethysmogram (PPG), neurotransmitters, blood glucose, blood pressure, temperature, hemoglobin, and blood flow as randomness source to generate cryptographic keys [67]–[69], [83]–[85]. These studies do not only employ PF as a randomness source but

| Solution Type | Employed Approach | Summary | Strength and Limitations | Ref. |
|---|---|---|---|---|
| Employing Physiological Features | Electrocardiogram (ECG) | • Generates an ECG-based random number by measuring the temporal and morphological ECG changes. The main idea is to use the obtained random number as a cryptographic key to authenticate the programmer to IMD.<br>• Data encryption is also proposed between IMD and programmer by generating a one-time pad from ECG signals. | **Strength**:<br>• The entropy of random numbers is high since randomness is obtained from a real physical event from the patient's body.<br>• There is no need for key distribution or key agreement since each party ends up with the same random number due to the same randomness source.<br>**Limitation**:<br>• Due to the noisy nature of ECG signals, IMD and programmer might <u>NOT</u> measure the same signal that affects the usability goals that requires re transmissions. | [64]–[66] |
| | Heart to Heart (H2H) and Interpulse Interval (IPI) | • Uses IPI data as an randomness source to generate cryptographic key that is used in both providing forward security and access control.<br>• Uses Neyman Person Lemma scheme for feature extraction from data instead of commonly used Hamming distance.<br>• Uses low-exponent RSA encryption, AES and hash computation during pairing to enhance security and less computation. | **Strength**:<br>• The entropy of randomness source is high enough to protect protocol against various attacks.<br>• It has a high noise tolerance due to employed feature extraction while generating random numbers.<br>**Limitation**:<br>• The scheme suffers from of false-positive and false-negative in case of noisy measurements.<br>• It is hard to measure IPI by biosensors. | [67], [68] |
| | Biometric (Blood pressure, blood glucose, Temperature, Hemoglobin, Blood flow) | • Proposes to use various biometrics to generate secure keys provide security of IMD data.<br>• They employed RC5 scheme as an pseudo-random number generator and used biometric features as an entropy source. | **Strength**:<br>• The use of biometrics to generate keys eliminate resource constrained computation and communication overhead.<br>**Limitation**:<br>• To reach an entropy desired to achieve secure randomness requires several biometric readings. The use of multiple biometrics measurements to establish efficient randomness with enough entropy might lead to a true rejection. | [69] |
| Employing Auxiliary | Vibration Key Exchange | • Proposes the use of motor-generated vibration for key exchange between the vibration generator and IMD.<br>• To achieve key exchange, vibration generator will send a secret key to IMD by modulating it into a vibration signal. | **Strength**:<br>• Due to the use of Ultra-low power (accelerometer), the two-step verification before starting to exchange key, the mechanism is resistant to battery drain attacks.<br>• Achieves a higher communication bit rate for key-exchange per second up to 20bps.<br>**Limitation**:<br>• Transmission rate is constrained by Inter-symbol interference (ISI) which reduces the reliable transmission. | [70], [71] |
| | Tatoo | • Uses ultraviolet-ink micropigmentation to encode a human readable key unto the patient's skin.<br>• Requires all devices that communicate with the IMD to have an ultraviolet light emitting diode (UV LED) and an input mechanism to enter the key to grant access to the IMD. | **Strength**:<br>• User can choose the character string they prefer, or a generated key can be used as a passcode.<br>**Limitation**:<br>• The key encoded on the patient body cannot be changed easily in situations where the key has been compromised and need to be renewed. | [72] |
| | Acoustic and Audio Key Exchange | • Proposes key exchange over acoustic wave. IMD and programmer achieve key exchange and authentication by using zero-power notification and zero-power human-sensible key exchange. | **Strength**:<br>• The scheme does not require additional computation or transmission to accomplish key exchange, therefore no additional energy cost for IMD.<br>**Limitation**:<br>• Found to be vulnerable to acoustic eavesdropping.<br>• The master key has to be stored securely on the programmer to achieve a maximum security. | [15], [73] [40] |

also use them as synchronized key agreement protocol. To do so, they benefit from the fact that IMD measures can also be measured by touching a patient's skin or observing it remotely with an external device (e.g., the programmer). They employ the measured PF independently and synchronously as a randomness source while ending up with the same symmetric key between IMD and the programmer device. Poon et al. [68] were the first who proposed a PF-based key generation method, claiming that the time between heartbeats generates a high amount of randomness to construct a secret key between parties. Although these studies mainly differ from each other in what PF they have used as a randomness source, they still have important differences in terms of how they employ PFs. Most of them omit to analyze the loss of entropy, which is an essential factor in using PFs for cryptographic key derivation. Another pitfall is generating cryptographic keys from noisy PF measurements and whether they take into consideration. A summary of these studies and the comparison according to mentioned pitfalls is presented in Table I.

### B. Employing Auxiliary Channels

An auxiliary channel uses vibration, visual, or acoustic channels for key agreements that are outside the typical communication channel between IMDs and programmer [15], [40], [70]–[73]. For example, Kim et al. [70] proposed, SecureVibe, a key agreement protocol for IMDs which uses a vibration-based auxiliary channel that is implemented with a simple vibration motor as a transmitter and an accelerometer sensor as a receiver. They use vibration states between on and off to represent 1 or 0 to transmit one bit per symbol. Schechter [72] proposed a visual auxiliary channel (e.g., ultra-violet visible tattoos) to retrieve pre-shared keys that are loaded IMD devices in manufacturing time. The programmer retrieves the pre-shared key under UV lights which is carved on the patient skin to authenticate itself to the IMD. Halperin et al. [15] proposed to use an audio channel to generate and transmit a shared key over the audio channel. These auxiliary channel-based key generation and agreement approaches are intrinsically secure due to their higher user perceptibility (dependent on user's perception and reaction) and close proximity. However, these

| Solution Type | Employed Approach | Summary | Strength and Limitations | Ref. |
|---|---|---|---|---|
| Employing External Devices | Cloaker | • Propose an externally worn cloaker device to mediate all communication between IMD and all pre-authorized parties. Cloaker provides confidentiality and authenticity by using symmetric key to encrypt and authenticate IMD and Cloaker. Communication between Cloaker and programmer is secure with public key cryptography. | **Strength**:<br>• Physician are allowed to access IMD even when Cloaker is lost, stolen, broken or runs out of batteries<br>• IMD ignores all communication from other parties when Cloaker is present while accepting all communication when the Cloaker is absent<br>**Limitation**:<br>• Patient is required to carry the Cloaker device as additional device among other devices a patient might have.<br>• IMD might be vulnerable when patient not to wear Cloaker device or device runs out of battery | [74] |
| | IMDGuard and RFID Guardian | • Proposes external wearable devices to authenticate prior to establishing communication with the IMD and operate in normal mode when patient is wearing the device and emergency mode otherwise. Key establishment scheme based on patient's ECG signals as the source of randomness to extract a symmetric key from ECG features. Uses symmetric key during IMD and programmer communication and uses public keys to authenticate the programmer and the guardian. | **Strength**:<br>• Resistant to spoofing attack by jamming the IMD's transmission of the challenge message.<br>•<br>**Limitation**:<br>• The guardian has to be worn by the patient at all times to avoid an adversary attack<br>• The guardian could be lost or stolen and expose IMD to attacks<br>• Vulnerable to Man-in-the-middle attacks when the attacker can physically measure real-time ECG signal of the patient. | [75], [76] [77] |
| | Shield | • Proposes a physical layer gateway solution between IMD and programmer to authorized endpoints and share keys. Uses novel radio designed to continuously listen and jam transmission.<br>• Shield and IMD share a communication channel that is not accessible to other parties, therefore ensuring reliable decoding of messages | **Strength**:<br>• Jamming of IMD messages causes eavesdroppers who are about 20cm away to experience about 50% error rate<br>• There is no modification requirement to the IMD with Shield integration<br>**Limitation**:<br>• Without the shield IMD listens and reply to all messages which open up for attack<br>• High powered adversary transmission within a few meters can interact with the IMD even when the Shield is present<br>• The data under jamming can still be extracted by the eavesdroppers using multiple-input multiple-output (MIMO)-based attack. | [78], [79] [80], [81] |
| | Physically Obfuscated Keys (POK) | • Proposes the use of integrated circuit to store secret key and adopt touch-to-access for emergency access.<br>• The use of One-time-programming enables one-time access, and the interface is disabled, and the secret is only accessible by microcontroller on the card.<br>• Achieve authentication and integrity through hash computation and token-based access. | **Strength**:<br>• Secret key stored on POK exist only when the chip is powered up and disappear after use to prevent illegal access and cannot be reconfigured.<br>• Use of biometric encryption of the temporary keys protect it from attackers<br>**Limitation**:<br>• Emergency access requires the use of a temporary key cached on the patient's IC card, which might not be accessible during emergency<br>• Loss of patient or physician's IC card could lead to attacks on the device | [64] |
| | Ultrasound Key Exchange | • Uses ultrasound reader as a body coupled communication to pair IMD and programmer to exchange keys.<br>• Randomly generated number and reader id is sent over ultrasound channel to establish a communication session with IMD. Both the IMD and the reader generates short term session key used to encrypt subsequent communication.<br>• Dual-processor architecture, medical processor for medical application and security processor for security protocol and communication. | **Strength**:<br>• Ultrasound channel is safer against eavesdropping within few feet<br>• To prevent battery DoS, ultrasound transducer converts incident waves into an electrical signal to wake up the security processor<br>**Limitation**:<br>• IMD assumes that all messages received from an Ultrasound channel are trusted since only trusted person can touch the patient for a long period of time<br>• Potential eavesdropping within a few feet from the patients | [82] |

studies slightly differ in usability, emergency situations, and key revocation. For instance, the tattoo-based mechanism is helpful for emergencies, but it has drawbacks in key revocation. A summary of studies and the comparison of them is presented in Table I.

## C. Employing External Device

Numerous studies have proposed using external devices to govern the security of IMDs including auditing, key management (e.g., generation, sharing, revocation, destruction), authentication, and access control [64], [74], [75], [77]–[82]. The main idea is to tackle the limitation of IMD resources (e.g., battery, memory, and processing power) while handling the security functions.

Denning et al. [74] is the first one who proposed the use of an external device called Cloaker. Cloaker is a wearable device that serves as a mediator between IMD and all other devices that are pre-authorized. The IMD is in a secure communication state if the patient wears the cloaker device. When patient wears the cloaker device and establishes communication chan-nel with IMD, IMD ignores all other communication. On the other hand, when cloaker device is not available, IMD accepts all communication which is a requirement for emergency situation. Confidentiality between IMD and cloaker is achieved by a symmetric cryptography while communication between cloaker and other devices (programmer) is secured by a public key cryptography.

In 2011, Xu et al. [75] proposed IMDGuard to establish a secure communication between IMD and other parties both in normal and emergency modes. Unlike Cloaker, IMDGuard serves as an authentication proxy with the help of patient's ECG signals. This is achieved when both the IMD and the IMDGuard extract patient's ECG signals simultaneously as keys without needing a prior shared key. This makes it resistant to key compromising attacks. IMDGuard also offers resistance against spoofing attacks by implementing a defensive jamming mechanism. It carries jamming challenges to prevent the intervention of communication by any other device rather than IMDGuard.

Gollakota et al. [78] also proposed an external device called

Shield to authorize endpoints and IMD. Unlike Cloaker and IMDGuard, Shield uses a novel approach by capturing all radio communication, whether it is authorized or unauthorized. It jams radio signals such that the attacker cannot decode authorized messages, and the IMD cannot decode unauthorized signals. Their work demonstrates a countermeasure to passive eavesdropping and a countermeasure to an active attacker by sending bogus commands to the IMD.

In 2018, Fu et al. [64] proposed Physiological Obfuscated Keys where the pre-shared secret keys are stored on Integrated Circuit (IC) card as an authentication and access control proxy. Moreover, unlike previous ones, POKS is highly integrated with the Hospital Authentication System (HAS) to establish a secure communication channel.

Recently, Siddiqi et al. [82] proposed SecureEcho, a device-pairing scheme using MHz-range Ultrasound for key exchange where the ultrasound device is placed against the skin of the patient (body-coupled communication channel) during key exchange. The close proximity requirement of SecureEcho ensures that IMD trusts all messages from Ultrasound devices.

A summary of previous security solutions by implementing external devices and comparison of them is presented in Table II.

## VI. Software/Firmware Update Security

Software/Firmware update is a commonly adapted process of remediating security vulnerabilities, fixing software bugs, and introducing new features post hardware or software deployment. This same process is not so common among implantable medical devices. Software/Firmware update can be deployed as a security solution within the IMD domain as shown in Fig. 1. However, ensuring the secure delivery of firmware to IMDs requires considerable effort due to resource limitations. IMDs have interconnected blocks of microprocessors, microcontrollers, transceivers, and their respective communication protocol stacks, which facilitate the communication among components. Different firmware is installed on these components to perform their respective functionalities. Evolving features of IMDs require frequent bugs and security fixes even after being implanted. Malfunctioning due to a failure of the critical components of IMD or adversarial attacks of software/firmware can cause significant harm to the patient.

Keeping IMDs up to date with bug fixes, security updates, and necessary feature updates is extremely important. There are several incidents where embedded systems were compromised through different attack vectors. In 2014, researchers were able to install a Trojan after presenting itself as mass storage to the embedded system using the USB device [86]. In 2010, a researcher accessed memory by manipulating network packets [86]. In 2011, a researcher showed an attack against the battery controller to cause overheating of the device [86]. Considering these threats, there is an urgent need to develop a secure software/firmware update mechanism for IMDs.

Herbold et al. [87] categorize software updates into alteration (modification of firmware), reverse engineering (firmware reconstruction), unauthorized firmware (unknown firmware source), and unauthorized device (install firmware on unauthorized device). To ensure a secure firmware update process, the originator of the firmware should be vetted and limited to the manufacturer of the IMD device. The firmware should be authentic. The firmware should be verified to ensure that the correct firmware is applied to the correct IMD. Lastly, there should be a mechanism to roll back to the previous firmware version in case of firmware update failure [88].

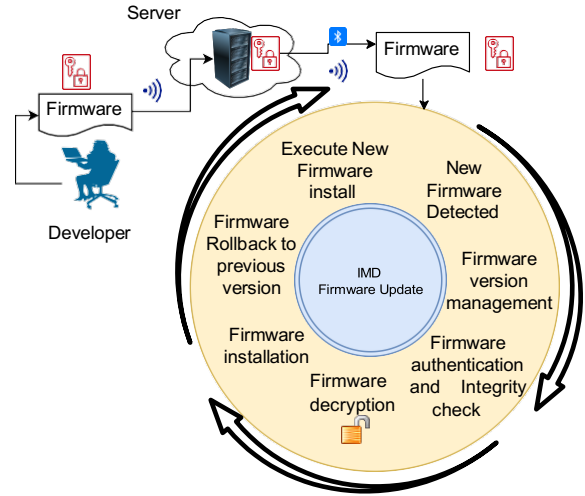

Fig. 5. Centralized Secure IMD Firmware Update

This section compares and contrasts the modes of secure firmware updates solutions offered for both Internet of Things (IoT) and IMD environments. Due to limited resources in these end devices, the offered security solution should also pay attention to the resource requirement. IMD firmware update is not heavily researched yet while writing this paper. Thus, we mainly focused on related firmware updates solutions in IoT devices to give insights about resource-aware secure software update solutions. Although IMD devices are different from IoT devices, we should note that they still share similar resource limitation and security concerns. Therefore, leveraging ideas from IoT domain will help researchers to improve the security posture of IMD devices.

One of the most common techniques that is employed during firmware update is Over-the-Air (OTA) update. OTA delivers firmware update using wireless connectivity. OTA is becoming the most popular method for firmware update due to its minimal human effort or interaction, cost effectiveness, convenience of use and the capability to deploy the firmware without a physical connection.

The OTA adaptation for IMD devices involves two devices: the target (IMD) and the controller (e.g., host computer, mobile phone, smartwatch, etc.,). The controller requests the firmware update and handles the transfer of the new firmware binary. The target is the end device that receives the updated new firmware binaries. A successful OTA firmware update requires two steps: the transportation of the firmware update package and replacing the old firmware with the new firmware package. The target and controller should ensure a secure firmware update through authentication, integrity check and successful rollback mechanism. OTA firmware update related studies is categorized under centralized and decentralized methods.

| Solution Type | Employed Approach | Summary | Strengths and Limitations | Ref. |
|---|---|---|---|---|
| Over-the-Air Firmware Update | Centralized | · Internet Accessible Raspberry PI: A Well adapted simple firmware update architecture by the research community to easily deploy firmware to end devices. However, the adaptation of this solution to IMD is limited since is not able to scale.<br>· General Packet Radio Service (GPRS): Global system of mobile communication has been researched to deliver firmware update to end devices. The challenge to this solution is load management and signal strength as the distance between the end device and the network increases. Researchers have employed geographical grouping system based a biased random key generation algorithm to group connected devices before firmware update.<br>· Firmware manager: Serves as intermediary between the server and end devices. Server sends the firmware to the manager and the manager fragments the firmware and distribute to the registered end devices. Due to limited bandwidth and lossy communication, employed fragmentation and logical layer acknowledgement increases reliability of this solution.<br>· A cloud-based solution: End devices periodically inquire from the cloud service for newer version of firmware. Manufacturers makes the new firmware available to the end device by uploading it to a cloud service. | **Usability and Mitigation to attacks**:<br>· Trust establishment between firmware manager and all end devices is required.<br>· The trust infrastructure presents a single point of failure.<br>**Integrity and Authenticity**:<br>· Firmware is delivered over a secure channel for example TLS.<br>· Employ symmetric encryption to calculate Message Authentication Code (MAC).<br>· Authentication is mostly achieved by previously distributed symmetric key. Secure key distribution is a challenge in IMD and IoT due to restricted resources. To overcome the key distribution problem, some solutions employ external device such as Physical Unclonable Function (PUF) of a device chip to generate their own one-time use key for authentication. | [89], [90] [86], [91], [88], [92] |
| | Distributed | · Blockchain: Uses distributed immutable ledgers, consensus algorithms and smart contracts to record and validate firmware update and related transactions. Permissioned blockchain uses six layered architecture to deliver firmware update. The layers include Application layer (apps), service layer (vendors), blockchain layer (smart contract), proxy layer, protocol layer (TLS, BLE) and the hardware layer(device).<br>· Seoyun et al implementation of blockchain use registration and retrieval nodes to register and download firmware images and manifest files.<br>· Lightweight mesh network protocol: This uses lightweight mesh network protocol over a peer-to-peer architecture to provide firmware update. The implementation uses a gateway to receive and transfer the firmware into the internal network, and intermediary hub between the gateway and end devices. | **Usability and Mitigation to attacks**:<br>· Resilient to most cyber-attacks due to the distributed nature of blockchain framework.<br>· Uses integrated Certificate Authority for membership management.<br>· The proprietary nature of LWMesh limits the usability, adaptability, and scalability of this solution.<br>**Integrity and Authenticity**:<br>· Integrity of firmware is ensured by the consensus protocol and hash values computed for the firmware image.<br>· Authenticity of the firmware update is ensured by the smart contract which implements the business logic. | [89], [90] [93] |
| In-field Physical Access Firmware Update | USB | · USB: Firmware binary are copied unto external USB drive and directly inserted into the USB port of the controller/programmer or the IMD to initiate the firmware update.<br>· Computer: Computer with access to the internet is used to connect to manufacturers network to initiate a firmware update. A data transfer cable such as mini-HDMI and USB cable is required to connect directly to the controller/programmer or the IMD to install the firmware. | **Usability and Mitigation to attacks**:<br>· USB based firmware update is only feasible for limited number of IMD solutions.<br>· Patients are scheduled to a hospital facility for the device to be updated.<br>**Integrity and Authenticity**:<br>· Access to the USB drive is managed by the device before allowing firmware update.<br>· Great effort is needed on the device to prevent a malicious action due to complexity of USB protocol.<br>· Since the user/patient would have to download updater to their computer, the assumption is that the user would first authenticate with the manufacturer's infrastructure before allowing the firmware update. | [94], [95] |
| | Debugging Interface | · Joint Test Action Group (JTAG): JTAG is an industry standard to access Test Access Port (TAP) for design verification, debugging, boundary scan testing and firmware delivery. It is considered to be quieter, faster and less expensive as compared to UART, USB or the bootloader when delivering firmware to flash or SRAM.<br>· Universal Asynchronous Receiver-Transmitter (UART): UART Interface is also industry standard debugging interface on mostly embedded devices allowing a serial connection between the device and the external resource. | **Usability and Mitigation to attacks**:<br>· ODM/OEM most often will blow out the JTAG fuse, obfuscate JTAG, or disabled JTAG to disallow the use of the JTAG to eliminate the attack interface. It is accessible through both software (e.g. bootloader) or hardware (e.g. adapter).<br>**Integrity and Authenticity**:<br>· ODM/OEM most often would implement firmware image integrity check before install. Signature verification, encryption and decryption mechanism are deployed before firmware image is transferred and installed. | [96], [97] [40] |

· **Centralized Approaches**:

A centralized firmware update mechanism is simply called as client-server model. Once a new software/firmware is ready, the firmware can be securely uploaded to a centralized location (such as cloud-based solution or server within manufacturer's infrastructure) which is made accessible to all authorized IMDs. Upon availability of the firmware on the server, the IMD detect the firmware and initiate the firmware installation process by first going through the version management process to ensure a safely and sound firmware update.

There are many studies that address security of OTA update [86], [89], [91], [92]. They mainly differ how they addressed the infrastructure related challenges. For instance, Hans et al. [89] addressed both reliability and security challenges on a mesh network protocol. Similar studies can give many insights for researcher in IMD domain while handling a firmware update employing a body area network [98]. Although authentication is mostly achieved by symmetric key which is previously distributed to end devices, the reliable distribution of update packets while ensuring its integrity become a significant challenge due to limited bandwidth and lossy communication. Some studies tackle this problem [86] and offers a integrity mechanism for fragmented packets via hash chaining to guarantee integrity of the fragmented firmware image.

- **Decentralized Approaches**: The client-server architecture has the drawback of having a Single Point of Failure (SPoF) when the update server is compromised. Besides, it does not allow involving a common and trusted auditing mechanism to check the validity of firmware. Thus, researchers offered to employ blockchain approach which requires a level of trust among all nodes participating in the network to ensure a secure firmware update. They attempt to replace a centralized solution with high availability and cryptographically secure to prevent unauthorized modification of data. Mehta et al. [99], proposed a blockchain based security for 5G-enabled Unmanned Aerial Vehicle (UAV) also known as drones to secure its applications. Banerjee et al. [100], used a blockchain to detect and self-heal compromised firmware that poses threat to IoT environment. This studies [101]–[103] propose a secure exchange of health data using blockchain. Although their framework presents how patients could securely connect, interact, and aggregate their health information which are managed by different stakeholders, it can guide researchers to implement a threat-sharing mechanism to employ in to for firmware update. A summary of examples of OTA firmware updates comparison is presented in Table III.

- **In-field Firmware Update**:
  The in-field firmware update requires a field engineer to be physically present and have a physical connection to the IMD. The firmware-update is accomplished through USB drive, Serial port, Ethernet port, Universal Asynchronous Receiver/Transmitter (UART) or Joint Test Action Group (JTAG). A commonly used technique to achieve in-field firmware-update employs the debugging interfaces of the chip such as JTAG and UART. However, these interfaces might not just enable updating IMD's firmware but also of tracing CPU instructions or reading memory sections. Thus, it suffers from significant attacks by allowing any universal programmer to change the firmware of these devices even after a secure update [40].

## VII. SENSORS AND ACTUATORS SECURITY SOLUTIONS

Sensors are mostly deployed in IMDs to monitor conditions based on inputs (e.g. temperature, coltage, current) to support therapy delivery to patients. Actuators on the other hand combines resources based on outputs (e.g signals and energy) to deliver therapy to patients. Sensors and actuators mostly work together to deliver therapy in IMD space (Close-loop system). Protecting sensors and actuators from adversary malicious attacks is also key to ensuring overall IMD security as depicted in Fig. 1.

Patients' physiological signals are measured using sensors to detect abnormal behavior while the actuators react to restore abnormal behavior or prevent the worst-case scenario.

For example, ICD can detect abnormal heartbeat and deliver a shock to the heart to return it to normal heartbeat [5], [105]. Internal and external sensors deployed within IMD environments communicate with the IMD either through a wireless interface or a physical interface for both external and internal sensors respectively. The sensors collect data and sends the data to IMD for a dynamic adjustment to settings without physicians. Before sensors can communicate securely with each other there has to be a trust establishment followed by a data communication which is mostly achieved by a cryptographic key [114]. Newaz et al categorized sensors in healthcare domain as Physiological sensors, biological sensors and Environmental sensors [48]. Sensors collect data and sends the aggregated data to the server. For example, a patient with insulin pump glucose monitoring system, the sensor measures the glucose level and sends the measurement to the insulin pump [105].

Both internal and external sensors and actuators can become a target of an adversary and expand the scope of IMD attack surface. Kune et al, researched into injecting electromagnetic interference signals into embedded sensors on ICD to change its reading which resulted in baseband and amplitude EMI attacks [104]. Ilias et al, researched into out of band acoustic signal injection attacks by transmitting higher amplitude signals into the sensors of Microelectromechanical Systems (MEMS) to report wrong values [116].

Security solutions to detect and prevent above mentioned sensors and actuators attacks are summarized in Table IV.

## VIII. DIFFERENCE BETWEEN IMD AND EMBEDDED SYSTEMS SECURITY SOLUTIONS

### A. Vulnerabilities

Common vulnerabilities identified in the IMD domain are similar to vulnerabilities found in embedded systems. For instance, IMDs suffer from code injection, malware, denial-of-service, the man in the middle, side-channel, and replay attacks. These attacks are also common among embedded systems. However, a disparity in IMD's attack vector is the proximity requirement that demands a close range to conduct some attacks compared to embedded systems.

### B. Security Solutions

The conventional security solutions to fulfill CIA triad goals can be employed for both IMD and embedded systems to mitigate the vulnerabilities mentioned above. However, since IMDs are implanted into the human body for a long time, there are challenges during the execution to achieve the same security goals compared to embedded systems. In addition, IMD has more severe resource constraints (such as battery life, memory, processing power) that prevent the IMD from adapting commonly deployed mitigations easily. For instance, a solution that demands costly communication and computation overhead is not desired in the IMD domain since it incurs the battery drain to shorten the battery life. On the other hand, the embedded systems could have their battery recharged or replaced, so it is acceptable to adopt costly computation to address the security goals.

| Solution Type | Employed Approach | Summary | Strength and Limitations | Ref. |
|---|---|---|---|---|
| Analog (Hardware-based) | Shielding | - Applies a conducting material to shield from electromagnetic radiation. | **Strength**:<br>- Producing EMI signal at 40bB with shield will force the adversary to transmit about $10^8$ times more power to have the same effect as without the shield.<br>- Commonly used to mitigate Electromagnetic Interference (EMI) attacks.<br>**Limitation**:<br>- Not fully protected from malicious electromagnetic since adversary with a powerful equipment may be able to emit signals that the shielding might not be able to prevent.<br>- How effective the shield is, mostly is dependent on the thickness of the shield. waves. Although | [5], [104] [105], [106] [107], [108] [109] |
| | Filtering | - Use to attenuate signals that are outside the baseband frequency thereby reducing the vulnerable frequency range of the sensor. | **Strength**:<br>- Commonly used to mitigate Electromagnetic Interference (EMI) and signal injection and waveform attacks.<br>**Limitation**:<br>- Do not have the detection capabilities | [5], [104] [105], [107] [109] |
| | Differential Comparators | - Uses a reference signal to remove common mode voltage interference in the sensor signal. | **Strength**:<br>- Differential signal measurement is in order when generated in a free air.<br>**Limitation**:<br>- Differential voltage drops significantly when not in free air and significant increased power brings it to order which an adversary can overcome with high power. | [5], [104] [105], [110] [107] |
| | Key Masking | - Randomized the secret key before each execution to prevent against side channel attack.<br>- Randomize the power profile by using flattening circuit or using another circuit as a bandpass filter. | **Strength**:<br>- Mitigate Differential Power Analysis (DPA)<br>**Limitation**:<br>- There is a lot of energy use overhead | [5], [111] [112] |
| Digital (Software-based) | Signal Contamination | - Estimating the EMI level when only the radiated signal are captured which helps to determine the defenses.<br>- Contamination level are set to help determine higher power pulses. | **Strength**:<br>- Commonly used to mitigate Electromagnetic Interference (EMI) and signal injection attacks.<br>**Limitation**:<br>- Vulnerable to adversary with strong emitter or adversary who may conduct baseband attacks | [105] [71] |
| | Adaptive Filtering | - Uses contaminated waveform to clean the received signal by estimating the RF-induced voltage.<br>- This is mostly adapted when contamination level of the signals exceed expected threshold to prevent improper actuation. | **Strength**:<br>- It can dynamically adjust the signals to determine a map which is used to translate waveform between components.<br>- Commonly used to mitigate waveform and signal injection attacks.<br>**Limitation**:<br>- The reaction time is dependent on the coefficient and might be a slower process.<br>- Vulnerable to adversary with strong emitter or adversary who may conduct baseband attacks | [105] |
| | Cardiac Probe | - Uses the results of actuation to determine if the system is under attack by comparing the sensor reading to an expected readings. For example, the direct connection with the cardiac tissue can be used to check a signal to distinguish between a measured and induced waveform. | **Strength**:<br>- The actuation process mostly cannot be observed by the adversary.<br>**Limitation**:<br>- Fault in the detection process might lead to the exposure of the secret key | [105], [113] |
| | Physiological Value-base Security (PVS) | - Uses PVS to establish a secure connection between two sensors | **Strength**:<br>- The process of both sensors generating PV is only done once and when there is a network connection problem and during network reconfiguration. No need to regenerate every time.<br>**Limitation**:<br>- There is resource demand and computational overhead and also the efficiency of this is dependent on PV which could be compromised by adversary. | [114] |
| | Dummy Sensor | - Uses sensor redundancy to detect signal injection attack. Since the second sensor doesn't has sensing part, micro-controller can measure its voltage output to adversary signal injection. | **Strength**:<br>- Detect signal injection attacks<br>**Limitation**:<br>- Duplicate sensors will increase the size of the device by twice the size of the sensor.<br>- cost of the device can become expensive<br>- Attacker surface is increased as new sensors are added | [106], [108] |
| | Sensor Modulation | - Uses duplicate sensors (active sensor and passive sensor) to detect signal injection by measuring the output from the active sensor and if there is no response there is a potential injection attack.<br>- Uses ON and OFF unpredictable state of the passive sensor to determine whether a voltage is measured ON and OFF state sensor, which will indicate an injection attack. | **Strength**:<br>- Detect signal injection attacks<br>**Limitation**:<br>- Detection is dependent on the active sensor<br>- Duplicate sensors will increase the size of the device by twice the size of the sensor.<br>- cost of the device can become expensive<br>- Attacker surface is increased as new sensors are added | [106], [115] [107] |

## C. Regulations and Privacy

IMD is a medical device that generates, accepts, uses, stores, manipulates and transfers patients' personal health information. Therefore, IMDs are subjected to Health Insurance Portability and Accountability Act (HIPAA) privacy rules and guidelines. Embedded systems, on the other hand, do not have a regulated standards, instead initiatives by EU and US [117], [118]. Moreover, the security of IMDs is strictly regulated by Food and Drug Administration (FDA) under the FDA Regulation of Medical Devices Act [119]. Thus, manufacturers must follow both pre-market and post-market policies [120]. However, embedded systems do not have such enforcing regulations to govern the security issues.

## IX. FUTURE RESEARCH ISSUES

Over the past few years, IMD security and privacy has caught the attention of researchers which has significantly improved security research in IMD domain. However, IMD domain still has a lot of open and known security research problems. In this section, we provide an overview of potential research path for future work on IMD security and privacy.

### A. Communication Protocols

Aram et al. proposed ZigBee the key and strong security features of ZigBee including cryptographic tools, cipher algorithm and fast and low-cost wireless communication [46]. The storage and power consumption requirements of ZigBee, possess a resource constraint on IMD and for that matter ZigBee adaptation into IMD domain. Further research into energy harvesting from surrounding energy sources and reducing the storage capacity will make its adoption into IMD space more desirable. BLE pairing mode with high level of security also consume a lot of power. Research into higher security level by encrypting authentication credentials, hiding UUID even at runtime, reconstructing the UUID at runtime, or onetime UUID and exploring BLE 5 security features with the help of energy harvesting or less power consumption will not only make BLE highly desirable in IMD domain but all miniaturized devices that need connectivity. Another possible future research work would be to design a new communication protocol dedicated to IMD and can communicate securely to physician network and data storage resources and it is protected from malicious attacks.

### B. Security Threats

Eduard et al. [105] proposed further research into overcoming the distance limitation of analogue attacks. To expand on that future work, researchers could attempt to conduct a remote attack leveraging patient's mobile devices which the patient will carry and is in close proximity to the IMD. Our current work also focusses on creating a threat modeling methodology using blockchain, threat intelligence and threat sharing to build and early warning system. As a future work, a threat monitoring device (e.g. mobile phone application) could be investigated whether the external device could monitor and alert potential threats on the IMD.

### C. Cryptographic Key Security

Public key cryptography (PKC) such as Rivest, Shamir and Adleman (RSA), Elliptic Curve Cryptography (ECC) and Attribute-Based Encryption (ABE) have been deployed in many domains to fulfil security requirements. [65], [121]–[125]. Due to resource constraints, the use of PKC in the IMD domain is not highly desirable. However, considering the recent improvements in manufacturing that make the processing power cheaper, there is a window to employ PKC in the IMD domain. Nevertheless, it still requires some further studies to address related challenges. Fortunately, the challenges are not completely unknown and previously studied. For instance, Nils et al. [79] did some experiments and showed that PKC could be used on small devices like IMDs without any hardware acceleration [79].

Michael et al. [126] addresses the possibilities of using short RSA secret exponents to increase performance.

Bos et al. [127] conducted a key length assessment of both RSA and ECC and compared the security level of different key sizes. These studies can shed light on further ones to address the employment of PKC in the IMD domain. In addition, researchers may deploy post-quantum techniques in IMD domain by investing the effect of algorithm key size changes and the use of one-time pad key counter quantum threats.

### D. Firmware Update

As of writing this paper, there were limited research into IMD software/firmware update. Due to IMD connectivity to hospital infrastructure and internet, the attack surface has also increased tremendously, and for that reason, IMD software/firmware update is a much-needed research to securely update IMD. Researchers can study the deployment of Blockchain-based firmware update and demonstrate the effectiveness of securely updating post surgical implant. Blockchain has been adapted in IoT domain to securely distribute firmware to end devices and hence could potentially be adapted into IMD.

### E. Sensors and Actuators Security

More advanced research is needed to ensure that sensors and actuators doesn't execute a malicious code and continue to operate as intended. To achieve this level of security, future advance research work could manufacture smart sensors and smart actuators capable of distinguishing between a legitimate command verses malicious command via machine learning. Machine learning has been adapted and researched into mostly all industry including IoT and healthcare systems. Machine learning could likewise be used in IMD space to advance sensors and actuators security posture.

## X. CONCLUSION

Implantable Medical Device is known to improve the quality of life of patients, restore self-confident and rekindle patients dreams which otherwise might not be possible and in some cases preserve lives. Advancement in IMD technologies and next generation of IMD are gradually becoming more connected and increasing in the communication and the way therapy is delivered to patients. This survey showed that, there are more IMD currently in the market and hopefully more IMD solutions in the near future. The review also presented commonly adapted communication protocols in IMD domain and also IMD collect and process data as patient uses the device. As IMD advance in technology and become more connected to the internet, there is a need to provide security solutions as counter measure to the threats that we reviewed. Some of the most pressing security solutions much needed to secure IMD which this survey reviewed were cryptograpic keys and software/firmware updates. In addition, we reviewed one of the difficult security problems with IMD which sensors and actuators and evaluated best solutions as a countermeasure to adversary's threats and attacks. While the benefits and the market acceptance of IMD have grown significantly, and the projected growth on revenue is projected to grow about 5% in couple of years, yet this era has also seen drastic increase in cyber-attacks on medical devices. Therefore, it is of optimal interest to proactively secure IMD devices and the privacy of patients. In conclusion, we believe this survey will have a greater impact in the IMD research community and motivate researchers and manufacturers alike, to develop mechanism that secure both legacy IMD devices in the market as well as new IMD design against adversary attacks.